\documentclass[A4paper,12pt]{article}
\usepackage[T1]{fontenc}
\usepackage[english]{babel}
\usepackage[cp1250]{inputenc}
\usepackage{epsfig}
\usepackage{amsfonts}
\usepackage{fancyhdr}
\usepackage{syntonly}
\usepackage{graphicx}

\begin{document}
\selectlanguage{english}

\begin{titlepage}
\begin{center}
\vspace*{3cm}

\begin{title}
\bold {\Huge Selecting the diffractive events \\ at the LHC energies
 }
\end{title}

\vspace{2cm}

\begin{author}
\Large K. FIA{\L}KOWSKI\footnote{e-mail address:
fialkowski@th.if.uj.edu.pl}

\end{author}

\vspace{1cm}

{\sl M. Smoluchowski Institute of Physics\\ Jagellonian University \\

30-059 Krak{\'o}w, ul.Reymonta 4, Poland}

\vspace{2cm}

\begin{abstract}
The PYTHIA 8 generator is used to estimate the percentage of the non-diffractive and
diffractive events at the LHC energies. It is shown that a simple condition of the absence
of charged hadrons in the central pseudorapidity region is sufficient to remove almost all
non diffractive events. This opens the way to investigate diffraction without
waiting for the future specialized detectors.
\end{abstract}

\end{center}

\vspace{2cm}



\end{titlepage}

\section{Introduction}
The first data on the multiplicity of hadrons produced at the LHC energies were
extensively analyzed and compared with lower energy data. An obvious difficulty
in such comparisons is the distinction between the non-diffractive (ND), single
diffractive (SD) and double diffractive (DD) events, for which all models predict
different energy dependence. In the data it is rather difficult to distinguish
unambiguously these classes of events. Thus the simplest way to proceed is to define
by some kinematic conditions the sample of events to be compared with models, and
then to impose the same conditions on the model predictions.
\par
An example of such a condition was introduced by the ALICE collaboration, where
the multiplicities in the pseudorapidity range from $-1$ to $1$ were measured for
the inelastic events having at least one charged hadron in this range \cite{ALICE}. Such a sample
is customarily denoted as INEL>0. It was shown \cite{KFRW} that the average multiplicities
for the INEL>0 sample are successfully described by the PYTHIA 8 generator
\cite{SMS}, \cite{SMS2}, \cite{TS}, if some parameters are properly tuned. However, it was also
noticed that this sample of events contains significant contributions from all the classes of events
mentioned above (ND, SD and DD). In the samples generated by PYTHIA 8  the proportion of these
contributions (ND:SD:DD) depends on the used version of MC and changes slowly with energy.
\par
This fact was regarded as disadvantageous, since a simple physical interpretation of the data
seems possible for the ND sample, and not for the unspecified mixture of the ND, SD and DD
events. However, one could note that the condition of at least one charged hadron in the
central pseudorapidity bin leaves the ND contribution almost intact, whereas a large percentage
of the SD and DD events is then removed.
\par
We use this observation to propose a simple way to construct a diffractive sample of events
from the LHC data without using specialized "diffractive" detectors, which are mostly in the
construction phase. The results from the PYTHIA 8 generator suggest that by requiring {\bf no} charged
hadrons in the central pseudorapidity bin we remove very effectively the ND contribution.
\par
In the next section we specify the versions of MC generators used and the generating conditions.
Then we present the results. We complete the note with some conclusions and outlook.

\section{Procedures and results}
We are using two recent versions of PYTHIA: 8.135 and 8.145. For the latter we check the
influence of the values of some tuning parameters. Specifically, we compare two choices of the
parameter values for the description of multiple scattering effects. The parameters define the
regularization of the (divergent) QCD cross section by introduction of the factor
$$F(p_T)=\frac{p_T^4}{(p_{T0}^2+p_T^2)^2}.$$
With an energy independent value of $p_{T0}$ the average multiplicity would increase too fast
with CM energy $E$. Thus a mild power-like dependence is assumed:
$$p_{T0}=p_{T0}^{ref} \left( \frac{E}{E^{ref}} \right) ^{\alpha}.$$
The default values of $E^{ref}$ and $\alpha$ are $1800$ GeV and $0.24$, respectively. The modified
values, for which the average multiplicities are better described, are $1000$ GeV and $0.30$.
In both cases the $p_{T0}^{ref}$ value is $0.2$ GeV/c.
We generate $100$k events for each energy, each class of events and each version of the MC generator.
Then we calculate the fraction of events which pass the condition of no charged hadrons in the central
pseudorapidity bin. The bins of one- and two units width are used.
\par
The results are summarized in Table 1. For transparency, the results for various versions of MC
generators are not shown separately, but the central value and the spread (in brackets) of percentages
is given. This spread should be a fair estimate of the model uncertainties.
\begin{table}[h]
\caption{The percentages of ND,  SD and DD events for various samples of data at three LHC energies from the PYTHIA 8 generator}
\begin{center}
\resizebox{0.95\textwidth}{!}{%
\begin{tabular}{||c|c|c|c|c||}
 \hline
 \hline
 \vspace{2pt}
       E (TeV)& Class of events& ND & SD & DD \\
\hline
\vspace{2pt}

          0.9 &Inclusive  &65.5(0.3) & 22.3(0.1)&12.2(0.1)\\
\hline
\vspace{2pt}
  0.9 &  $N=0$ in $\Delta \eta =1$ & 21.5(4.5)&50.7(2.2) &28.0(1.5) \\
\hline
\vspace{2pt}
     0.9 &  $N=0$ in $\Delta \eta = 2$ &5.0(1.8) & 62.1(1.0)&33.0(0.8) \\
\hline
\vspace{2pt}
     2.0 &  Inclusive &66.0(0.2) & 21.2(0.2)&12.7(0.2) \\
     \hline
\vspace{2pt}
     2.0 &  $N=0$ in $\Delta \eta =1$ &19.8(3.6) &50.2(2.6)& 29.3(1.8) \\
     \hline
\vspace{2pt}
     2.0 &  $N=0$ in $\Delta \eta = 2$ &4.3(1.6) & 60.8(1.3)&35.4(0.8)\\
     \hline
\vspace{2pt}
     7.0 &  Inclusive &67.5(0.2) & 19.4(0.3)&13.0(0.2) \\
     \hline
\vspace{2pt}
     7.0 &  $N=0$ in $\Delta \eta = 1$ &19.1(5.2) & 48.5(3.7)&32.2(1.6) \\
\hline
\vspace{2pt}
     7.0 &  $N=0$ in $\Delta \eta = 2$ &3.3(1.4) & 57.3(3.6)&39.5(2.3) \\
\hline \hline
\end{tabular}
}
\end{center}
\end{table}
\par
We see that already the condition of no hadrons in the central pseudorapidity bin of the length of one unit
removes the great part of ND events. By requiring no hadrons in central two units of $\eta$ we get an almost pure
diffractive sample: less than one event in $20$ is non-diffractive. Obviously, some diffractive events are also rejected,
but the loss is usually less than a half of the sample.

\section{Conclusions and outlook}
We have checked that in the PYTHIA 8 generator a simple condition to accept only the events without charged hadrons
in the central two units of pseudorapidity removes almost all the non-diffractive events. Thus an almost pure
diffractive sample of events may be easily selected.
\par
Obviously, this conclusion may not be valid for other generators, In particular, if a model predicts a long tail
of the rapidity gap distribution in non-diffractive sample, the results may be quite different. However, it seems
that the PYTHIA 8 describes various LHC data well enough to justify the confidence that our results apply to the data.
\par
The possibility to extract the diffractive contribution from the existing data without waiting for the dedicated diffractive
experiments seems quite attractive. An example of a possible use of such procedure may be the investigation of the charge
asymmetry for $W^{+/-}$ boson production. It is well known that the excess of valence $u$ quarks in protons results in an
excess of positive leptons from the $W$ decays \cite{ATLAS}, \cite{CMS}. Such an excess should not occur for so-called
double Pomeron exchange ("central diffraction"), since the $u/d$ quarks (antiquarks) are perfectly balanced in the Pomeron.
For the single Pomeron exchange, i.e. the SD and DD samples, the asymmetry should be about half of that observed for the ND events.
However, there are models in which soft quark/gluon exchange in the final state change these expectations. Thus testing such
predictions seems worthwhile \cite{GB}. With our condition one may do it easily. Other applications are certainly possible.

\section{Acknowledgments} 
I would like to thank Andrzej Bia{\l}as, Andrzej Kotanski, Micha{\l} Prasza{\l}owicz and Romuald Wit for helpful remarks.

\end{document}